\pdfoutput=1
\documentclass[12pt,a4paper]{article}

\usepackage{ifthen} 
\newboolean{pdflatex}
\setboolean{pdflatex}{true} 

\newboolean{articletitles}
\setboolean{articletitles}{true} 

\newboolean{uprightparticles}
\setboolean{uprightparticles}{false} 

\def\paperauthors{LHCb collaboration} 
\def\paperasciititle{Template for writing LHCb papers} 
\def\papertitle{Measurement of the effective leptonic weak mixing angle} 
\def\paperkeywords{{High Energy Physics}, {LHCb}} 
\def\papercopyright{\the\year\ CERN for the benefit of the LHCb collaboration} 
\def\paperlicence{CC BY 4.0 licence}
\def\paperlicenceurl{https://creativecommons.org/licenses/by/4.0/}

\newif\ifEnableSectionTOCLinks
\EnableSectionTOCLinksfalse 

\usepackage[top=1in, bottom=1.25in, left=1in, right=1in]{geometry}

\usepackage{microtype}
\usepackage{lineno}  
\usepackage{xspace} 
\usepackage{caption} 

\usepackage{graphicx}  
\usepackage{color}
\usepackage{colortbl}
\graphicspath{{./figs/}} 

\usepackage{amsmath} 
\usepackage{amssymb}
\usepackage{amsfonts}
\usepackage{upgreek} 

\newcommand*\patchAmsMathEnvironmentForLineno[1]{%
\expandafter\let\csname old#1\expandafter\endcsname\csname #1\endcsname
\expandafter\let\csname oldend#1\expandafter\endcsname\csname
end#1\endcsname
 \renewenvironment{#1}%
   {\linenomath\csname old#1\endcsname}%
   {\csname oldend#1\endcsname\endlinenomath}%
}
\newcommand*\patchBothAmsMathEnvironmentsForLineno[1]{%
  \patchAmsMathEnvironmentForLineno{#1}%
  \patchAmsMathEnvironmentForLineno{#1*}%
}
\AtBeginDocument{%
\patchBothAmsMathEnvironmentsForLineno{equation}%
\patchBothAmsMathEnvironmentsForLineno{align}%
\patchBothAmsMathEnvironmentsForLineno{flalign}%
\patchBothAmsMathEnvironmentsForLineno{alignat}%
\patchBothAmsMathEnvironmentsForLineno{gather}%
\patchBothAmsMathEnvironmentsForLineno{multline}%
\patchBothAmsMathEnvironmentsForLineno{eqnarray}%
}

\usepackage{hyperxmp}

\usepackage[pdftex,
            pdfauthor={\paperauthors},
            pdftitle={\paperasciititle},
            pdfkeywords={\paperkeywords},
            pdfcopyright={Copyright (C) \papercopyright},
            pdflicenseurl={\paperlicenceurl}]{hyperref}

\usepackage[bottom,flushmargin,hang,multiple]{footmisc}

\usepackage[all]{hypcap} 

\usepackage{xspace} 
\usepackage{upgreek}

\def\lhcb   {\mbox{LHCb}\xspace}

\def\MagUp {\mbox{\em Mag\kern -0.05em Up}\xspace}

\ifthenelse{\boolean{uprightparticles}}%
{

 \def\Pmu         {\ensuremath{\upmu}\xspace}

 \def\Ppsi        {\ensuremath{\uppsi}\xspace}

 \def\PDelta      {\ensuremath{\Delta}\xspace}                 
 \def\PXi         {\ensuremath{\Xi}\xspace}                 
 \def\PLambda     {\ensuremath{\Lambda}\xspace}                 
 \def\PSigma      {\ensuremath{\Sigma}\xspace}                 
 \def\POmega      {\ensuremath{\Omega}\xspace}                 
 \def\PUpsilon    {\ensuremath{\Upsilon}\xspace}
 \let\oldPi\Pi
 \def\PPi         {\ensuremath{\oldPi}\xspace}

 \def\PB      {\ensuremath{\mathrm{B}}\xspace}                 
 \def\PD      {\ensuremath{\mathrm{D}}\xspace}                 
 \def\PJ      {\ensuremath{\mathrm{J}}\xspace}                 
 \def\PK      {\ensuremath{\mathrm{K}}\xspace}                 
 \def\PW      {\ensuremath{\mathrm{W}}\xspace}                 
 \def\PZ      {\ensuremath{\mathrm{Z}}\xspace}                 
 \def\Ps      {\ensuremath{\mathrm{s}}\xspace}

 \def\thebaroffset{0.0em}
}
{

 \def\Pmu         {\ensuremath{\mu}\xspace}

 \def\Ppsi        {\ensuremath{\psi}\xspace}                 
                  
 \mathchardef\PDelta="7101
 \mathchardef\PXi="7104
 \mathchardef\PLambda="7103
 \mathchardef\PSigma="7106
 \mathchardef\POmega="710A
 \mathchardef\PUpsilon="7107
 \mathchardef\PPi="7105
 \def\PB      {\ensuremath{B}\xspace}                 
 \def\PD      {\ensuremath{D}\xspace}                 
 \def\PJ      {\ensuremath{J}\xspace}                 
 \def\PK      {\ensuremath{K}\xspace}                 
 \def\PW      {\ensuremath{W}\xspace}                 
 \def\PZ      {\ensuremath{Z}\xspace}                 
 \def\Ps      {\ensuremath{s}\xspace}

 \def\thebaroffset{0.18em}
}
\newcommand{\offsetoverline}[2][\thebaroffset]{\kern #1\overline{\kern -#1 #2}}%

\makeatletter
\ifcase \@ptsize \relax
  \newcommand{\miniscule}{\@setfontsize\miniscule{4}{5}}
\or
  \newcommand{\miniscule}{\@setfontsize\miniscule{5}{6}}
\or
  \newcommand{\miniscule}{\@setfontsize\miniscule{5}{6}}
\fi
\makeatother

\DeclareRobustCommand{\optbar}[1]{\shortstack{{\miniscule (\rule[.5ex]{1.25em}{.18mm})}
  \\ [-.7ex] $#1$}}

\def\mumu       {{\ensuremath{\Pmu^+\Pmu^-}}\xspace}

\def\squark    {{\ensuremath{\Ps}}\xspace}

\def\KorKbar {\kern \thebaroffset\optbar{\kern -\thebaroffset \PK}{}\xspace}

\def\D       {{\ensuremath{\PD}}\xspace}

\def\DorDbar {\kern \thebaroffset\optbar{\kern -\thebaroffset \PD}\xspace}

\def\Dp      {{\ensuremath{\D^+}}\xspace}
\def\Dm      {{\ensuremath{\D^-}}\xspace}

\def\DpDm    {\ensuremath{\Dp {\kern -0.16em \Dm}}\xspace}

\def\B       {{\ensuremath{\PB}}\xspace}

\def\BorBbar {\kern \thebaroffset\optbar{\kern -\thebaroffset \PB}\xspace}

\def\Bd      {{\ensuremath{\B^0}}\xspace}

\def\BdorBdbar {\kern \thebaroffset\optbar{\kern -\thebaroffset \Bd}\xspace}

\def\Bs      {{\ensuremath{\B^0_\squark}}\xspace}

\def\BsorBsbar {\kern \thebaroffset\optbar{\kern -\thebaroffset \Bs}\xspace}

\def\jpsi     {{\ensuremath{{\PJ\mskip -3mu/\mskip -2mu\Ppsi}}}\xspace}

\def\Upsilonres  {{\ensuremath{\PUpsilon}}\xspace}
\def\Y#1S{\ensuremath{\PUpsilon{(#1S)}}\xspace}

\def\LorLbar     {\kern \thebaroffset\optbar{\kern -\thebaroffset \PLambda}\xspace}

\def\to                 {\ensuremath{\rightarrow}\xspace}

\def\AT#1     {\ensuremath{A_{\mathrm{T}}^{#1}}\xspace}           

\def\C#1      {\ensuremath{\mathcal{C}_{#1}}\xspace}                       
\def\Cp#1     {\ensuremath{\mathcal{C}_{#1}^{'}}\xspace}                    
\def\Ceff#1   {\ensuremath{\mathcal{C}_{#1}^{\mathrm{(eff)}}}\xspace}        
\def\Cpeff#1  {\ensuremath{\mathcal{C}_{#1}^{'\mathrm{(eff)}}}\xspace}       
\def\Ope#1    {\ensuremath{\mathcal{O}_{#1}}\xspace}                       
\def\Opep#1   {\ensuremath{\mathcal{O}_{#1}^{'}}\xspace}                    

\newcommand{\aunit}[1]{\ensuremath{\text{\,#1}}}       

\newcommand{\tev}{\aunit{Te\kern -0.1em V}\xspace}
\newcommand{\gev}{\aunit{Ge\kern -0.1em V}\xspace}
\newcommand{\mev}{\aunit{Me\kern -0.1em V}\xspace}
\newcommand{\kev}{\aunit{ke\kern -0.1em V}\xspace}
\newcommand{\ev}{\aunit{e\kern -0.1em V}\xspace}
 
\newcommand{\mevc}{\ensuremath{\aunit{Me\kern -0.1em V\!/}c}\xspace}
\newcommand{\gevc}{\ensuremath{\aunit{Ge\kern -0.1em V\!/}c}\xspace}
\newcommand{\mevcc}{\ensuremath{\aunit{Me\kern -0.1em V\!/}c^2}\xspace}
\newcommand{\gevcc}{\ensuremath{\aunit{Ge\kern -0.1em V\!/}c^2}\xspace}

\def\gsim{{~\raise.15em\hbox{$>$}\kern-.85em
          \lower.35em\hbox{$\sim$}~}\xspace}
\def\lsim{{~\raise.15em\hbox{$<$}\kern-.85em
          \lower.35em\hbox{$\sim$}~}\xspace}

\def\pt         {\ensuremath{p_{\mathrm{T}}}\xspace}

\def\ptot       {\ensuremath{p}\xspace}

\def\evtgen     {\mbox{\textsc{EvtGen}}\xspace}

\def\geant      {\mbox{\textsc{Geant4}}\xspace}

\def\photos     {\mbox{\textsc{Photos}}\xspace}

\def\pythia     {\mbox{\textsc{Pythia}}\xspace}

\def\tell1  {TELL1\xspace}
\def\ukl1   {UKL1\xspace}

\newcommand{\lhcborcid}[1]{\href{https://orcid.org/#1}{\hspace*{0.1em}\raisebox{-0.45ex}{\includegraphics[width=1em]{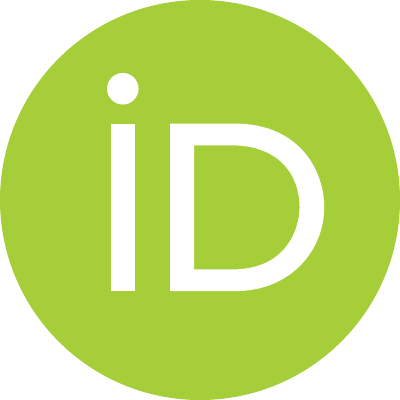}}}}

\hypersetup{
  colorlinks   = true, 
  urlcolor     = blue, 
  linkcolor    = blue, 
  citecolor    = red   
}

\ifEnableSectionTOCLinks
    \usepackage[explicit]{titlesec} 
    
    \let\oldcontentsline\contentsline
    \renewcommand

    \titleformat{\section}{\normalfont\Large\bf}{\hyperlink{tocsection.\thesection}{{\thesection} \parbox[t]{\dimexpr\textwidth-1pc}{#1}}}{1pc}{}

    \titleformat{\subsection}{\normalfont\bf}{\hyperlink{tocsubsection.\thesubsection}{{\thesubsection} \parbox[t]{\dimexpr\textwidth-1pc}{#1}}}{1pc}{}

    \titleformat{name=\section,numberless}[display]{}{}{0pt}{\normalfont\Huge\bfseries #1}
\fi

\usepackage{cite} 
\usepackage{mciteplus}

\usepackage{longtable} 

\begin{document}

\newcommand{\stwlept}{\ensuremath{\sin^2\theta_{\rm eff}^\ell}\xspace}
\newcommand{\afb}{\ensuremath{A_{\rm FB}}\xspace}
\newcommand{\deltaeta}{\ensuremath{|\Delta \eta|}\xspace}
\newcommand{\signeddeltaeta}{\ensuremath{\Delta \eta}\xspace}
\newcommand{\powhegbox}{{\sc{Powheg-Box}}\xspace}
\newcommand{\powhegewnlo}{{\sc{Powheg}}-ewnlo\xspace}
\newcommand{\powhegewlo}{{\sc{Powheg}}-ewlo\xspace}
\newcommand{\powhegplain}{{\sc{Powheg}}-plain\xspace}
\newcommand{\dyturbo}{{\sc{DYTurbo}}\xspace}

\newcommand{\NonLinearityShift}{\ensuremath{+2.0 \times 10^{-5}}\xspace}
\newcommand{\MomentumSmearingChiSq}{1564}
\newcommand{\MomentumSmearingNDOF}{919}
\newcommand{\FinalResultValue}{0.23147}
\newcommand{\FinalResultStatErr}{0.00044}
\newcommand{\FinalResultExpErr}{0.00005}
\newcommand{\FinalResultTheoryErr}{0.00023}

\renewcommand{\thefootnote}{\fnsymbol{footnote}}
\setcounter{footnote}{1}


\begin{titlepage}
\pagenumbering{roman}

\vspace*{-1.5cm}
\centerline{\large EUROPEAN ORGANIZATION FOR NUCLEAR RESEARCH (CERN)}
\vspace*{1.5cm}
\noindent
\begin{tabular*}{\linewidth}{lc@{\extracolsep{\fill}}r@{\extracolsep{0pt}}}
\ifthenelse{\boolean{pdflatex}}
{\vspace*{-1.5cm}\mbox{\!\!\!\includegraphics[width=.14\textwidth]{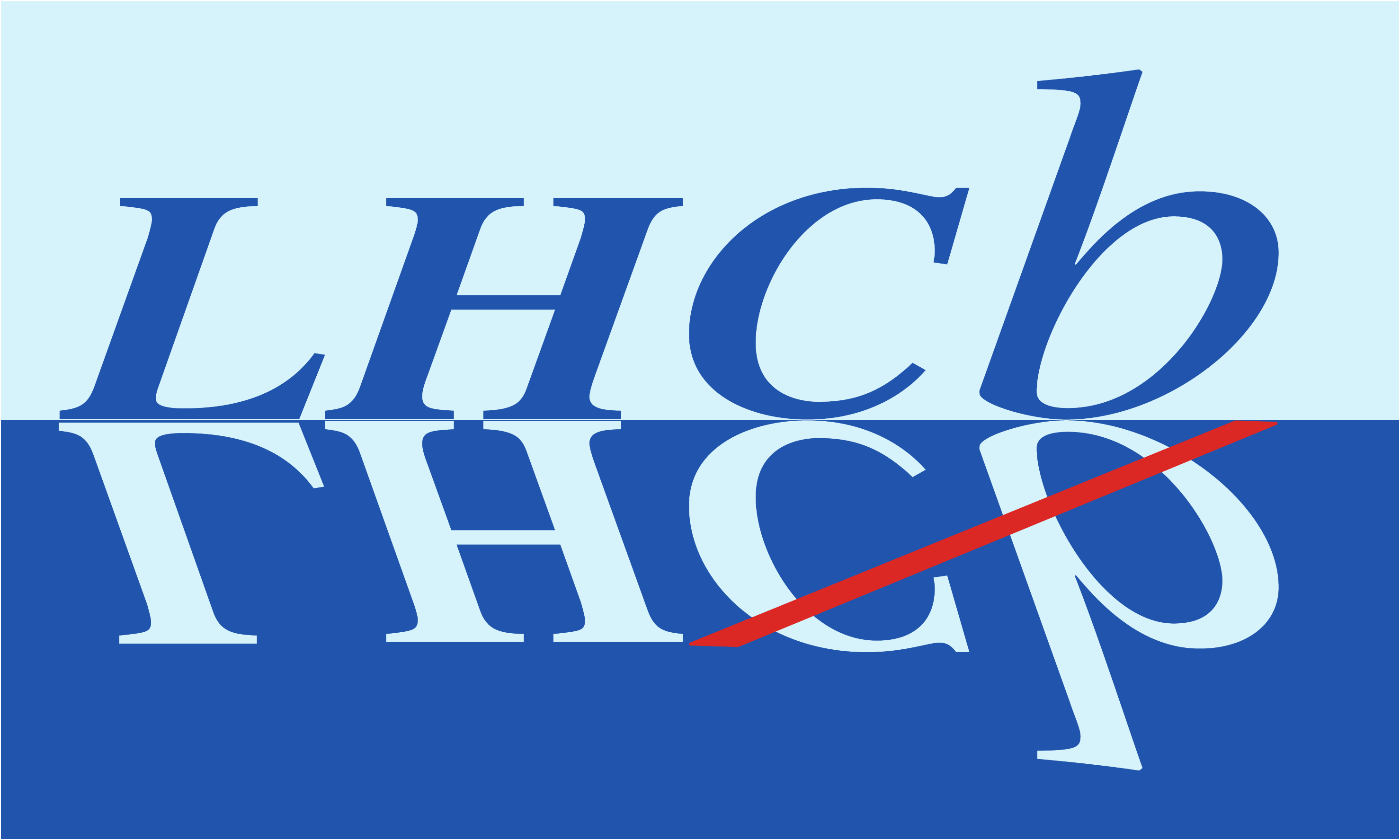}} & &}%
{\vspace*{-1.2cm}\mbox{\!\!\!\includegraphics[width=.12\textwidth]{figs/lhcb-logo.eps}} & &}%
\\
 & & CERN-EP-2024-230 \\  
 & & LHCb-PAPER-2024-028 \\  
 & & December 6, 2024 \\ 
 & & \\
\end{tabular*}

\vspace*{1.0cm}

{\normalfont\bfseries\boldmath\huge
\begin{center}
  \papertitle 
\end{center}
}

\vspace*{1cm}

\begin{center}
\paperauthors\footnote{Authors are listed at the end of this paper.}
\end{center}

\vspace{\fill}

\begin{abstract}
\noindent
Using $pp$ collision data at $\sqrt{s}=13$~TeV, recorded by the LHCb experiment between 2016 and 2018 and corresponding to an integrated luminosity of $5.4$~fb$^{-1}$, the forward-backward asymmetry in the $pp \to Z/\gamma^{*} \to \mu^+\mu^-$ process is measured. The measurement is carried out in ten intervals of the difference between the muon pseudorapidities, within a fiducial region covering dimuon masses between $66$ and $116$~GeV, muon pseudorapidities between $2.0$ and $4.5$ and muon transverse momenta above $20$~GeV.
These forward-backward asymmetries are compared with predictions, at next-to-leading order in the strong and electroweak couplings. The measured effective leptonic weak mixing angle is
\begin{equation*}
\stwlept = \FinalResultValue \pm \FinalResultStatErr \pm \FinalResultExpErr \pm \FinalResultTheoryErr,
\end{equation*}
where the first uncertainty is statistical, the second arises from systematic uncertainties associated with the asymmetry measurement, and the third arises from uncertainties in the fit model used to extract \stwlept from the asymmetry measurement. This result is based on an arithmetic average of results using the CT18, MSHT20, and NNPDF31 parameterisations of the proton internal structure, and is consistent with previous measurements and with predictions from the global electroweak fit.
\end{abstract}

\vspace*{2.0cm}

\begin{center}
  Published in
  JHEP 12 (2024) 026
\end{center}

\vspace{\fill}

{\footnotesize 
\centerline{\copyright~\papercopyright. \href{\paperlicenceurl}{\paperlicence}.}}
\vspace*{2mm}

\end{titlepage}


\newpage
\setcounter{page}{2}
\mbox{~}
%
%
%
%


\renewcommand{\thefootnote}{\arabic{footnote}}
\setcounter{footnote}{0}


\cleardoublepage


\pagestyle{plain} 
\setcounter{page}{1}
\pagenumbering{arabic}

\section{Introduction}
\label{sec:Introduction}
 
The weak mixing angle $\theta_{W}$ is one of the fundamental parameters of the Standard Model; at lowest order it relates the values of the U(1) and SU(2) gauge couplings. 
Consequently, it controls the couplings of the \PZ boson: the tree-level vector coupling to an elementary fermion of charge $Q$ and third weak-isospin component $T_3$ is $T_3 - 2Q\sin^2\theta_W$.
Higher-order corrections to the couplings are then included by defining an effective angle, which for leptons can be written via
\begin{equation}
  \stwlept \equiv \kappa_{\rm lept} \sin^2\theta_W,
\end{equation}
where the factor $\kappa_{\rm lept}$ contains both universal and flavour-specific terms~\cite{LEPWeakMixing}.
The weak mixing angle is scale dependent; 
we define \stwlept to be evaluated at a renormalisation scale equal to the mass of the $Z$ boson. 
The value of \stwlept can be predicted by global electroweak fits~\cite{Haller:2018nnx,PhysRevD.106.033003}, and a comparison of these predictions to direct measurements is sensitive to possible corrections involving fields beyond those present in the Standard Model. This article reports a measurement of \stwlept using data collected with the LHCb detector at the Large Hadron Collider (LHC).

The two most precise measurements of \stwlept are from the forward-backward asymmetry in $e^+e^- \to \PZ \to b\bar{b}$ processes at LEP~\cite{LEPWeakMixing} and the leptonic coupling asymmetry at the SLD experiment~\cite{SLDWeakMixing}. These two results are in tension at the level of $3.2$ standard deviations. Additional measurements have also been combined by the LEP experiments~\cite{LEPWeakMixing}. Measurements at hadron colliders have also been reported by the  ATLAS~\cite{ATLASWeakMixing}, CMS~\cite{CMSWeakMixing} and LHCb~\cite{LHCb-PAPER-2015-039} experiments at the LHC, and by the CDF and D0 experiments at the Tevatron~\cite{TevatronWeakMixing}.

At hadron colliders \stwlept can be determined 
from
$\PZ \to \ell^+\ell^-$ production,\footnote{For brevity we use $Z$ to refer to the physical process including amplitudes with $Z$ and virtual photon propagators.} where $\ell$ is an electron or muon.
The differential cross-section follows~\cite{CollinsSoperFrame1977,Korner:1990im}
\begin{equation}
  \label{eq:diffcross}
  \frac{\mathrm{d}\sigma}{\mathrm{d}\cos\theta^\ast} \propto 1 + \cos^2\theta^\ast + \alpha\cos\theta^{\ast},
\end{equation}
where $\theta^\ast$ is the polar angle in a suitable frame. 
In the Collins--Soper frame~\cite{CollinsSoperFrame1977}, $\theta^\ast$ can be calculated from variables in the laboratory frame via
\begin{equation}
\label{eq:ast}
\cos\theta^{\ast} = \frac{2(P_1^+P_2^- - P_1^-P_2^+)}{\sqrt{m_{\ell\ell}^2(m_{\ell\ell}^2+p_{\rm T,\ell\ell}^2)}} \frac{p_{z,\ell\ell}}{|p_{z,\ell\ell}|},
\end{equation}
where $p_{\rm T,\ell\ell}$, $p_{z,\ell\ell}$ and $m_{\ell\ell}$ are the transverse momentum, longitudinal momentum and mass of the dilepton system, respectively. The $P_i^\pm\equiv\frac{1}{\sqrt{2}}(E_i\pm p_{z,i})$ terms are calculated from the energies ($E$) and longitudinal momenta ($p_{z}$) of the lepton and antilepton, which are labelled with $i$ values of 1 and 2, respectively. 
The final factor in Eq.~\ref{eq:ast}, corresponding to the sign of $p_{z,\ell\ell}$, is required in proton-proton ($pp$) collisions given the symmetry of the initial state. 

The coefficient $\alpha$ in Eq.~\ref{eq:diffcross} arises through terms involving products of vector and axial-vector couplings and can therefore be directly related to the weak mixing angle. In addition, since the relevant term in Eq.~\ref{eq:diffcross} is linear in $\cos\theta^{\ast}$ it also directly causes a forward-backward asymmetry in $\PZ \to \ell^+\ell^-$ production; measurements of this asymmetry can then be used to determine \stwlept. The forward-backward asymmetry is typically defined as 
\begin{equation}
  \afb \equiv \frac{\sigma_{\rm F} - \sigma_{\rm B}}{\sigma_{\rm F} + \sigma_{\rm B}},
\end{equation}
where $\sigma_{\rm F,B}$ are the cross-sections integrated over the ranges $0< \cos\theta^{\ast} < 1$ (forward, F) and $-1 < \cos\theta^{\ast} < 0$ (backward, B). Since events with the largest values of $|\cos\theta^{\ast}|$ are most sensitive to the linear term in Eq.~\ref{eq:diffcross}, these events also provide the greatest sensitivity to the weak mixing angle. Therefore, in some previous analyses~\cite{ATLASWeakMixing,CMSWeakMixing}, 
the weak mixing angle has been extracted from an angular analysis or by measuring \afb using a per-event weighting that depends on $\cos\theta^\ast$~\cite{afbweighting}. In this paper we follow a related approach, by considering \afb in intervals of the absolute difference between the pseudorapidities of the two muons produced in the \PZ boson decay, \deltaeta. Since $\cos\theta^{\ast} \sim \tanh (\signeddeltaeta / 2)$~\cite{phistar} this choice enables us to separate the events with the greatest sensitivity to the weak mixing angle. In
simulation this binning improves sensitivity to the weak mixing angle by 14\,\% when compared to an approach with no binning in \deltaeta. For simplicity, following this binning choice, we also define `forward' and `backward' labels based on the sign of the difference in pseudorapidity of the muons. This is of negligible consequence: the assigned `forward' or `backward' label is different with this choice to that using the Collins--Soper angle for only one candidate \PZ decay in the analysis reported in this article. In summary, this analysis measures
\begin{equation}
  \label{eq:afb_eventcounts}
  \afb \equiv \frac{N(\eta^- > \eta^+) -N(\eta^- < \eta^+)}{N(\eta^- > \eta^+) +N(\eta^- < \eta^+)},
\end{equation}
as a function of \deltaeta, where $N$ denotes a yield of events passing the requirements in parentheses corrected for detector effects, and $\eta^-$ and $\eta^+$ are the pseudorapidities of the negatively and positively charged leptons, respectively. 

The unique angular coverage of the LHCb detector is well suited for this measurement. Important theoretical uncertainties in measurements of the weak mixing angle at hadron colliders typically arise from existing knowledge of the proton's internal structure. At the LHC, these uncertainties are smallest in the forward region~\cite{Azzi:2019yne,LHCb-PUB-2018-013}, such that the measurement from LHCb provides important complementary information to measurements performed at ATLAS and CMS. 

This analysis uses $pp$ collision data at a center-of-mass energy of $13$~TeV, recorded with the LHCb detector during 2016, 2017 and 2018, and corresponding to an integrated luminosity of $5.4$~fb$^{-1}$. The analysis is carried out in two parts. In the first stage \afb is measured in ten intervals of \deltaeta up to $\deltaeta = 2.5$, using $Z\to \mu^+\mu^-$ decays.
The asymmetries are measured in the fiducial region corresponding to dimuon masses in the range $66 < M < 116$~GeV, and with individual muon pseudorapidities in the range $2.0 < \eta < 4.5$ and transverse momenta $\pt > 20$~GeV.\footnote{Throughout this paper we use natural units, where $c=1$. We also define $\pt$, $\eta$ and the dimuon invariant mass based on the stable final-state particles (commonly referred to as being measured at bare level).}
The second stage of the analysis compares the measurement with theoretical templates  to determine \stwlept.
In order to prevent human bias, the analysis has been carried out by introducing an unknown offset in the \stwlept value until the analysis methodology was finalised.

\section{Dataset}
\label{sec:dataset}

The \lhcb detector~\cite{LHCb-DP-2008-001,LHCb-DP-2014-002} is a single-arm forward spectrometer, which covers the \mbox{pseudorapidity} range $2<\eta <5$.
The detector includes a high-precision tracking system consisting of a silicon-strip vertex detector surrounding the $pp$ interaction region~\cite{LHCb-DP-2014-001}, a large-area silicon-strip detector (the TT) located upstream of a dipole magnet with a bending power of about $4{\mathrm{\,T\,m}}$, and three stations of silicon-strip detectors and straw
drift tubes~\cite{LHCb-DP-2017-001} placed downstream of the magnet.
Roughly half of the data were recorded with the magnet in each of the two polarity configurations.
The tracking system provides a measurement of the momentum, \ptot, of charged particles with a relative uncertainty that varies from 0.5\% at low momentum to 1.0\% at 200~GeV.
The minimum distance of a track to a primary $pp$ collision vertex (PV) is referred to as the impact parameter (IP), which is precisely determined by the vertex detector.
Different types of charged hadrons are distinguished using information from two ring-imaging Cherenkov detectors~\cite{LHCb-DP-2012-003}.
Photons, electrons and hadrons are identified by a calorimeter system consisting of scintillating-pad and preshower detectors, an electromagnetic calorimeter and a hadronic calorimeter. 
Muons are identified by a system composed of alternating layers of iron and multiwire proportional chambers~\cite{LHCb-DP-2012-002}.

This analysis uses events selected by the hardware trigger based on the presence of a muon with a high transverse momentum.
The software trigger performs a full event reconstruction, and this analysis selects events based on the presence
of high-transverse-momentum muon candidates~\cite{LHCb-DP-2019-001}.

Simulation is required to model and correct for the effects of the detection efficiency and resolution, and backgrounds. 
In the simulation, $pp$ collisions are generated using \pythia~\cite{Sjostrand:2007gs,*Sjostrand:2006za} with a specific \lhcb configuration~\cite{LHCb-PROC-2010-056}.
Decays of heavy particles such as weak bosons, and top quarks, are modelled directly with \pythia, while decays of lighter particles are described by \evtgen~\cite{Lange:2001uf}, in which final-state radiation is generated using \photos~\cite{davidson2015photos}.
The interaction of the generated particles with the detector, and its response, are implemented using the \geant toolkit~\cite{Allison:2006ve, *Agostinelli:2002hh} as described in
  Ref.~\cite{LHCb-PROC-2011-006}. 

Candidate $Z\to\mu^+\mu^-$ decays are formed from combinations of oppositely charged and positively identified muons, with $\pt > 20$~GeV and $2 < \eta < 4.5$,
and with dimuon invariant masses between $66$ and $116$~GeV.
After this initial selection the level of background is already very low, but several additional requirements are imposed to further improve the sample purity, with minimal reduction in the 
detection efficiency for the signal.
Both muons are required to have a small IP with respect to the relevant PV in order to suppress background
from heavy-flavour-hadron decays, and their corresponding track fits must have $\chi^2$ values below $2.5$ to suppress hadronic backgrounds. 
The sum of the transverse momenta of particles within $(\Delta\eta)^2 + (\Delta\phi)^2 < 0.4^2$ of each muon must be less than $40$~GeV (where $\phi$ denotes the azimuthal angle). This requirement suppresses hadronic backgrounds since they typically have increased activity close to the muons.
In order to precisely define the trigger efficiency, each candidate is required to have at least one muon that satisfies the requirements of the hardware and software triggers. After all these requirements around $860\,000$ events are selected. 

The decays $\jpsi \to \mu^+\mu^-$ and $\Upsilonres(1S) \to \mu^+\mu^-$ are used to calibrate the detection efficiency (discussed in detail in Sec.~\ref{sec:simulation_corrections}) and the muon momentum measurement.
Candidates for both decays are formed from combinations of oppositely charged tracks identified as muons with $\pt > 3$~GeV. 
The $\jpsi \to \mu^+\mu^-$ candidates are required to form a vertex that is significantly displaced from any PV; this implies that the signal originates from decays of $b$ hadrons.

Two calibrations are applied to the muon momenta in the data. 
The first is to correct for gradual variations of the momentum scale with time, known to be at the $\mathcal{O}(10^{-4})$ level~\cite{LHCb-DP-2023-003}.
Multiplicative correction factors are determined from the observed variation of the $\Upsilonres(1S)\to\mumu$ peak position 
in intervals of the data-taking period.
The second correction addresses charge-dependent curvature biases using the pseudomass method~\cite{Barter:2021npd,LHCb-DP-2023-001}.
The pseudomass is an estimate of the mass of two-particle final states, in which the magnitude of one of the momenta is ignored.
Considering the decay $Z\to\mu^+\mu^-$ we define the two pseudomasses:
\begin{equation} 
\mathcal{M}^{\pm} \equiv \sqrt{2p^{\pm}\pt^{\pm}\frac{p^{\mp}}{\pt^{\mp}}\left(1 - \cos\vartheta\right)}, 
\end{equation}
where $p^+$ and $p^-$ denote the magnitudes of the $\mu^+$ and $\mu^-$ momenta (and similarly for the transverse momenta $\pt^{\pm}$), and $\vartheta$ is the opening angle between the two muons.
Effectively, the pseudomass estimates the dimuon invariant mass under the assumption that $Z$ bosons are produced with transverse momenta much smaller than their mass. For a perfectly aligned detector, we expect to a very good approximation that the $\mathcal{M}^+$ and $\mathcal{M}^-$ distributions should agree. 
However, unlike the dimuon invariant mass, in which charge-dependent curvature biases strongly cancel, the pseudomasses have first-order sensitivity to these biases, thereby allowing these effects to be easily determined. 
In intervals of $\eta$, $\phi$, year and magnet polarity, a simultaneous fit of the positive and negative pseudomass is performed
to find the pseudomass asymmetry. This is then directly translated to provide corrections for biases in measurements of the charge-over-momentum, $q/p$. The difference between the $q/p$ biases found in data and simulation is then applied as a correction to data; this approach eliminates a small bias due to the presence of vector and axial-vector couplings in the physics process. It is shown comprehensively in Ref.~\cite{LHCb-DP-2023-001} that this effect is both small, and has minimal dependence on the value of \stwlept assumed in the simulation.

\section{Corrections to the simulation and background modelling}
\label{sec:simulation_corrections}

The simulation is used to model the detection efficiency and backgrounds, and subsequently to correct the data for these contributions.
Corrections to the simulation are required to improve the accuracy of this modelling, with systematic uncertainties then associated with these corrections. 

Some of the effects contributing to the momentum resolution are underestimated in the simulation; smearing of the momenta in the simulation is therefore required. The approach taken here closely follows that in the LHCb measurement of the \PW boson mass~\cite{LHCb-PAPER-2021-024}, using selected $\jpsi$, $\Upsilon(1S)$ and \PZ-boson events. The information provided by each of these three resonances is complementary, due to the different average momenta of the muons produced. The impact of this smearing on the final result is negligible.

The detection efficiency for the $Z\to\mu^+\mu^-$ signal is roughly 85\%, with the main contributors to the inefficiency being the trigger, track reconstruction and muon identification. 
Corrections are applied to the simulation in order to improve the accuracy with which the detection efficiency is modelled.
The trigger efficiency is measured in both data and simulation using a combination of $Z\to\mu^+\mu^-$ candidates, which provide constraints at high \pt, and $\Upsilonres(1S)\to\mu^+\mu^-$ candidates, which provide constraints at lower \pt. 
Candidates are required to have one muon that satisfies the requirements of the trigger; the other muon is therefore not required to trigger the recording of the event. 
In intervals of the direction of the other muon, the efficiency is estimated by the fraction of these candidates in which both muons satisfy the trigger requirements. 
Nine and four intervals are simultaneously used in $\eta$ and $\phi$, respectively, and the efficiency estimates are further divided into intervals of \pt. 
In each angular interval, the \pt dependence of the efficiency in the simulation is modelled with an error function, while the ratio of the efficiency in data to that in simulation is modelled with a linear function. 
These functions are used to assign a weight to each simulated event, depending on whether one or both muons satisfy the trigger requirement. 

The muon identification efficiency is determined in a similar way, using only \mbox{$Z\to\mu^+\mu^-$} candidates. 
A dedicated sample of \mbox{$Z\to\mu^+\mu^-$} candidates is selected with one muon allowed to fail the muon identification requirements, while the other must match the standard requirement. 
The fraction of these candidates in which both muons satisfy the requirements provides an estimate of the efficiency. 
By comparing these estimates for the data and simulation, weights are assigned to the simulated events based on parametric functions of \pt, determined in intervals of $\eta$ and $\phi$.

The track reconstruction efficiency is also determined with a dedicated sample of \mbox{$Z\to\mu^+\mu^-$} candidates in which one muon is reconstructed using only information from the TT and the muon subdetectors.
The fraction of events in which the muon is also found by the standard track-reconstruction algorithms~\cite{LHCb-DP-2013-002} provides an estimate of the efficiency. 
Corresponding weights are assigned to the simulated events. Unlike the trigger and muon identification, 
the tracking efficiency corrections have no significant \pt dependence for the high \pt muons studied.

The backgrounds 
in
the $Z\to\mu^+\mu^-$ samples are modelled using simulation. 
The total background fraction, within the kinematic region in which \afb is measured, is only $2\times 10^{-3}$.
Most backgrounds have steeply falling mass distributions, and are therefore relatively small in the region $66 < M < 116$~GeV.
The two largest background contributions arise from $Z\to\tau^+\tau^-$ decays and from the decays of heavy-flavour hadrons. Both these contribute to the sample with  fractions of around $5\times 10^{-4}$. Contributions from rarer processes are also considered, including weak-boson pair production, top-quark pair production, single-top-quark production, the production of $W$ bosons associated with hadrons misidentified as muons, and events with two hadrons misidentified as muons.

\begin{figure}[!ht!]\centering
\includegraphics[width=0.7\linewidth]{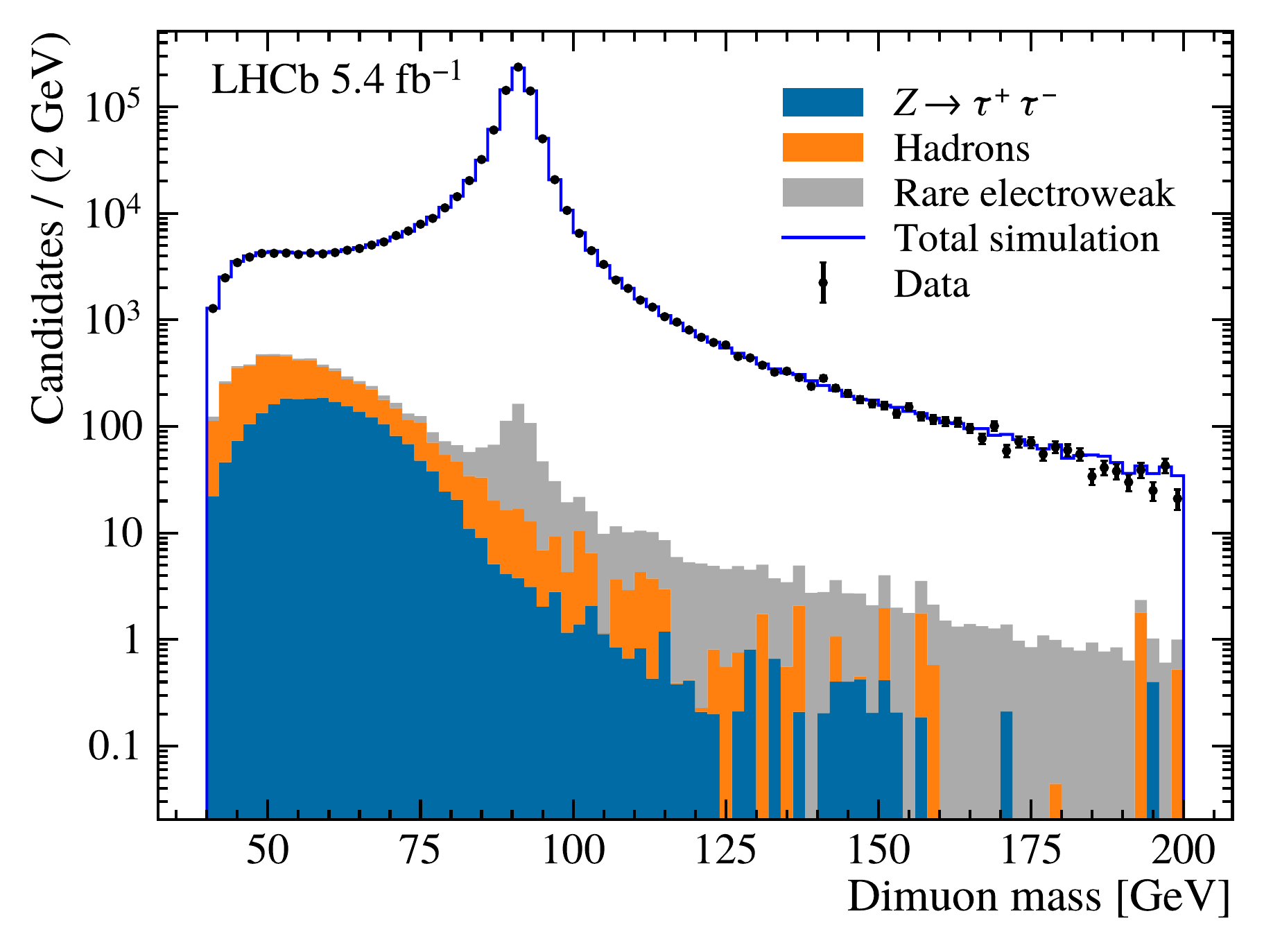}
\includegraphics[width=0.7\linewidth]{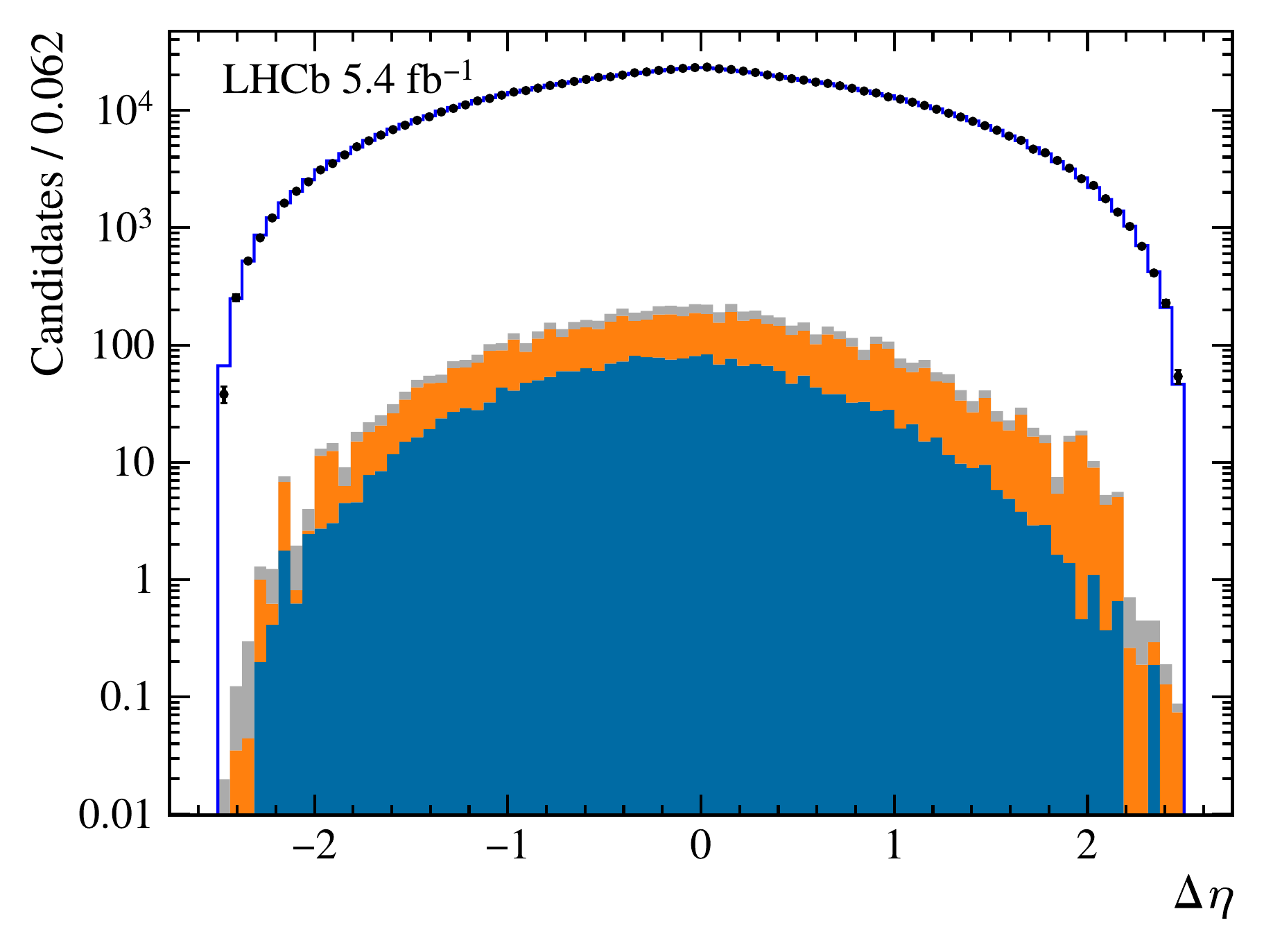}
\caption{Distributions of (top) dimuon invariant mass and (bottom) \signeddeltaeta for the selected signal candidates compared to simulation. The dark blue solid line corresponds to the sum of the expected signal and background contributions.}
\label{fig:simulation_control_plots}
\end{figure}

Figure~\ref{fig:simulation_control_plots} compares the dimuon invariant mass and \signeddeltaeta distributions of the selected candidates in data to simulation that includes both signal and background contributions. 
Both distributions are well described by the simulation, and it can be seen that the background level is extremely low.

\section{Measurement of the forward-backward asymmetry}
\label{sec:measurement}

\begin{figure}[!ht!]\centering
  \includegraphics[width=0.7\linewidth]{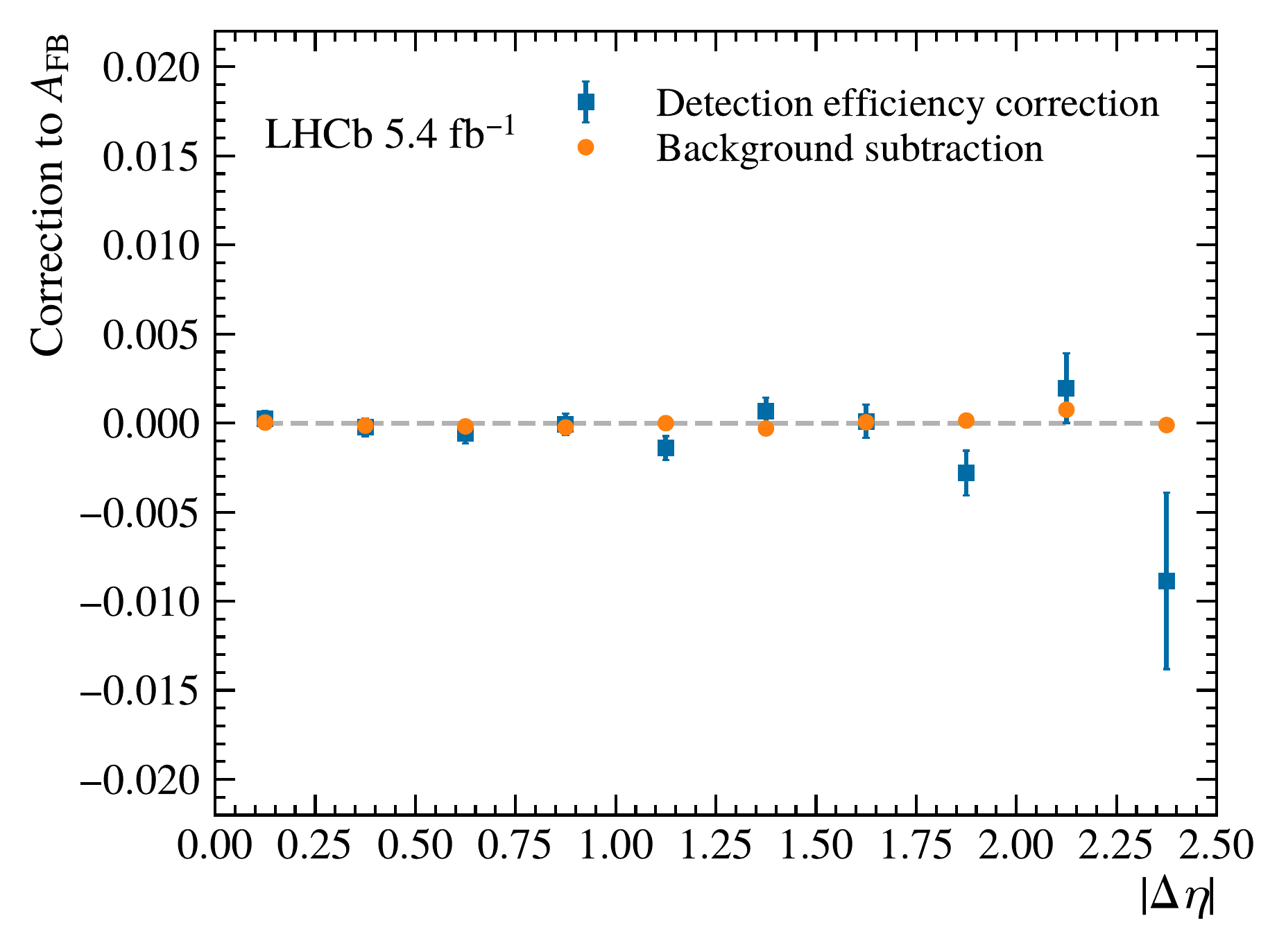}
  \caption{Effects of the detection efficiency correction and background subtraction on the measured \afb in ten intervals of $|\Delta\eta|$, shown in terms of the shift they introduce ($\Delta\afb$).
    All intervals are defined within the volume $|\Delta\eta| < 2.5$ and $66 < M < 116$~GeV.
    The error bars on the efficiency correction represent statistical uncertainties only, while the statistical uncertainties on the background subtraction are negligible.
  }
  \label{fig:afb_corrections}
\end{figure}

The measurement of the forward-backward asymmetry proceeds 
by 
measuring the forward and backward yields in ten intervals of \deltaeta and finding \afb following Eq.~\ref{eq:afb_eventcounts}. Corrections are necessary to account for the presence of background and detector effects such as 
inefficiencies.
Figure~\ref{fig:afb_corrections} shows the numerical effect of these two corrections.
The background is modelled as described above, with the background yields directly subtracted from the forward and backward yields. This correction is seen to have a very small effect on the measured \afb values.
The correction for detector effects is typically at the $\mathcal{O}(10^{-4})$ level. Systematic effects of a few $\times 10^{-3}$ are seen in some intervals, though these are subject to larger statistical uncertainties due to the finite simulation
sample sizes. We discuss the correction for detector effects in more detail below.

\subsection{Detector effects}

The measured value of \afb in each \deltaeta interval is corrected using a term determined in simulation, 
\mbox{$\afb^\text{true} - \afb^\text{reco}$}.
The value of $\afb^\text{true}$ is defined using truth information and all events in the fiducial acceptance, while $\afb^\text{reco}$ is defined using reconstruction-level information and only the events which pass the analysis selection requirements. 
This therefore corrects for: 
\begin{enumerate}
\item events missed due to detection and selection inefficiencies;
\item events missed due to (net) migration across boundaries in \pt, $\eta$ and dimuon mass, and moving in and out of the acceptance;
\item events reconstructed in the wrong interval of \deltaeta. 
\end{enumerate}
This last effect is negligible since the detector resolution on \deltaeta is excellent. An additional cross-check is performed incorporating this specific effect as a separate correction, which finds a negligible change in the final results. The overall correction could depend on the size of the weak mixing angle assumed in simulation. 
Both the correction and the final result are stable with respect to large changes in the assumed value of the weak mixing angle. 

\subsection{Systematic uncertainties}

\begin{figure}
    \centering
    \includegraphics[width=0.7\textwidth]{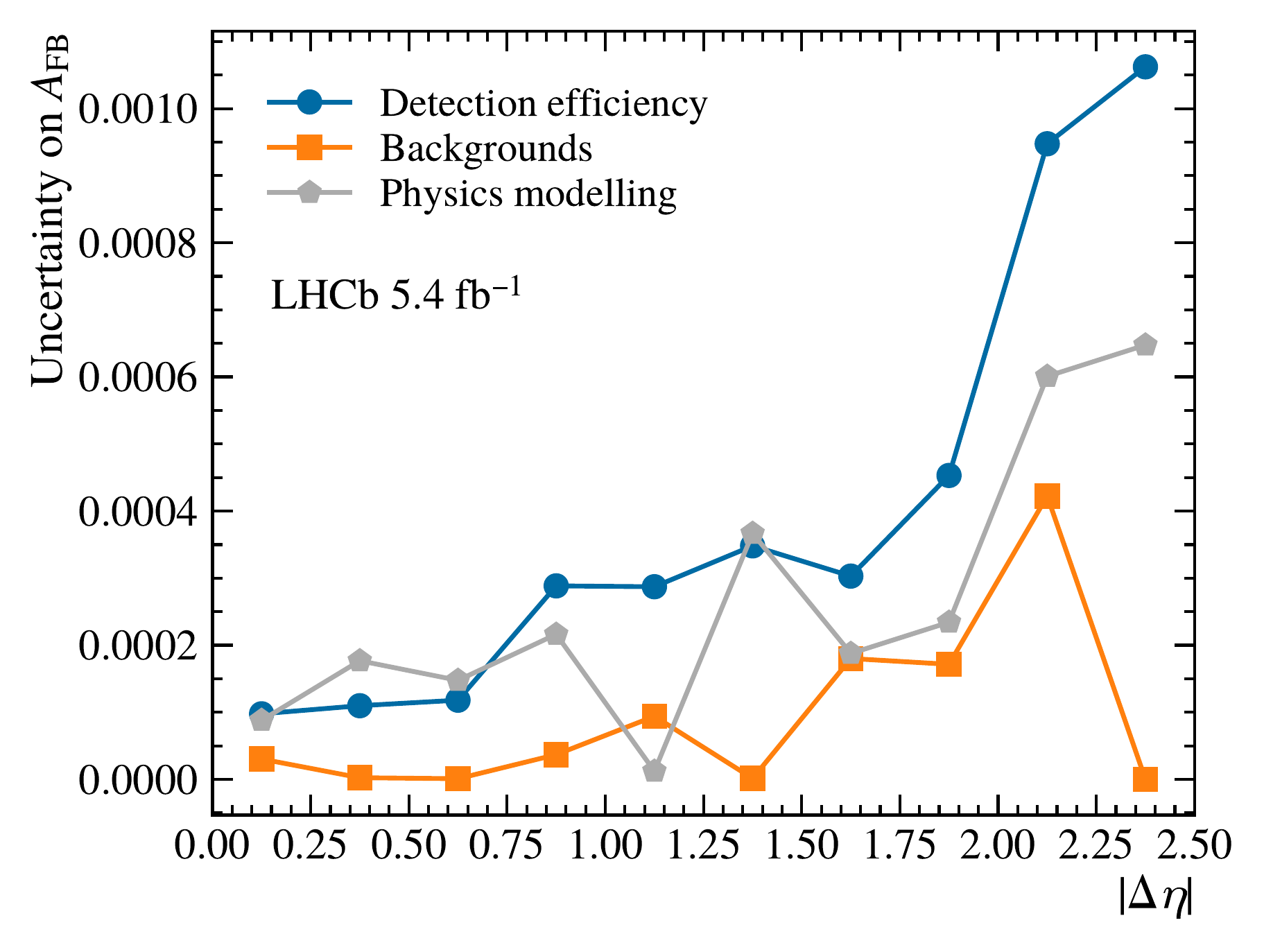}
    \caption{Systematic uncertainties on the \afb measurement in \deltaeta intervals.}
    \label{fig:afb_systematics}
\end{figure}

Figure~\ref{fig:afb_systematics} shows the sizes of the systematic uncertainties on \afb in the \deltaeta intervals.
These uncertainties are defined as follows.
\begin{description}
    \item[Detection efficiency:] The statistical uncertainties on the trigger, muon identification and tracking efficiency corrections are propagated by randomly varying the estimated efficiencies within their uncertainties and then redetermining the parameters of the \pt-dependent functions. This is then propagated through the measurement of \afb. For each efficiency factor, the uncertainty is defined by the root mean square of the resulting distribution of \afb values after the random variations. Discrete variations in the efficiency correction method are also considered. Tighter and looser requirements on the dimuon invariant mass and on the muon selection criteria are considered as an additional source of uncertainty in the determination of the muon efficiencies. 
    Since the efficiencies are studied in intervals covering detector regions, the number of intervals is varied. The variation that induces the largest change in \afb is then used to set an uncertainty. In addition, three alternative functional forms for the \pt-dependence of the efficiency corrections are also considered in the same way. Each contribution 
    is then combined in quadrature to set an overall detection efficiency uncertainty.
    \item[Backgrounds:] The cross-section assumed for the heavy-flavour-hadron background is varied up and down by 50\% with the resulting shifts in the measured \afb is defined as the associated uncertainty. The contribution from $Z\to\tau^+\tau^-$ decays occurs at a similar rate to the heavy-flavour-hadron background, but is known to far better precision, and consequently the uncertainty associated with this process is negligible. No uncertainty is assigned for other, smaller backgrounds. 
    \item[Physics modelling:] Weights are assigned to the  signal events such that the kinematic distributions match the predictions of the \dyturbo program~\cite{Camarda:2019zyx}, which has a higher formal accuracy than \pythia~8. The cross-section is predicted using \dyturbo in intervals of boson \pt, mass and rapidity, with logarithms in $\pt/M$ resummed to next-to-next-to-leading order (NNLO), while the angular coefficients are predicted at NLO in the strong coupling. These weights primarily affect the \afb measurement via changes in the detection efficiency correction. The shift in the \afb measurement sets the uncertainty.
\end{description}
The statistical uncertainties on the pseudomass calibrations and the momentum smearing are propagated through the \afb measurement, but their effect is found to be negligible, which is expected since the measurement only has a single wide interval in mass. The total uncertainty is found by combining the contributions from these different sources in quadrature.

\subsection{Results}

Table~\ref{tab:afb_results} and Fig.~\ref{fig:afb_and_stw_fit} report the measured \afb values in the ten intervals of \deltaeta. 
There are no correlations between the statistical uncertainties. 
The correlation matrix of the systematic uncertainties is presented in Table~\ref{tab:numerical_results_correlations}.

\begin{table}[t]\centering 
\caption{\label{tab:afb_results}Results of the \afb measurement. The first uncertainty is statistical and the second is systematic.}
\begin{tabular}{ccc}
Interval number & Interval&$A_{\rm FB}$\\
\hline
0&$0.00 < \deltaeta \leq 0.25$ & $0.0036\pm0.0025\pm0.0001$\\
1&$0.25 < \deltaeta \leq 0.50$ & $0.0204\pm0.0027\pm0.0002$\\
2&$0.50 < \deltaeta \leq 0.75 $ & $0.0303\pm0.0028\pm0.0002$\\
3&$0.75 < \deltaeta \leq 1.00$ & $0.0406\pm0.0031\pm0.0003$\\
4&$1.00 < \deltaeta \leq 1.25$ & $0.0466\pm0.0034\pm0.0002$\\
5&$1.25 < \deltaeta \leq 1.50$ & $0.0528\pm0.0039\pm0.0004$\\
6&$1.50 < \deltaeta \leq 1.75$ & $0.0622\pm0.0047\pm0.0003$\\
7&$1.75 < \deltaeta \leq 2.00$ & $0.0545\pm0.0060\pm0.0004$\\
8&$2.00 < \deltaeta \leq 2.25$ & $0.0603\pm0.0088\pm0.0010$\\
9&$2.25 < \deltaeta \leq 2.50$ & $0.0622\pm0.0190\pm0.0008$\\
\end{tabular}
\end{table}

\begin{table}[!hb!]\centering
\caption{\label{tab:numerical_results_correlations}Correlation coefficients for the experimental systematic uncertainties
  on the \afb measurement in ten intervals of $|\Delta \eta|$, with the interval numbers indicated as defined in Table~\ref{tab:afb_results}.}
\begin{tabular}{c|cccccccccc}
&0&1&2&3&4&5&6&7&8&9\\
\hline
0&$+1.00$&$-0.57$&$-0.66$&$-0.62$&$-0.16$&$-0.66$&$-0.83$&$-0.90$&$+0.31$&$+0.76$\\
1&$-0.57$&$+1.00$&$+0.92$&$+0.63$&$-0.09$&$+0.91$&$+0.45$&$+0.33$&$-0.68$&$-0.50$\\
2&$-0.66$&$+0.92$&$+1.00$&$+0.44$&$+0.22$&$+0.77$&$+0.41$&$+0.37$&$-0.82$&$-0.40$\\
3&$-0.62$&$+0.63$&$+0.44$&$+1.00$&$-0.62$&$+0.86$&$+0.60$&$+0.59$&$-0.15$&$-0.89$\\
4&$-0.16$&$-0.09$&$+0.22$&$-0.62$&$+1.00$&$-0.33$&$+0.08$&$+0.12$&$-0.18$&$+0.47$\\
5&$-0.66$&$+0.91$&$+0.77$&$+0.86$&$-0.33$&$+1.00$&$+0.63$&$+0.52$&$-0.47$&$-0.74$\\
6&$-0.83$&$+0.45$&$+0.41$&$+0.60$&$+0.08$&$+0.63$&$+1.00$&$+0.93$&$+0.11$&$-0.67$\\
7&$-0.90$&$+0.33$&$+0.37$&$+0.59$&$+0.12$&$+0.52$&$+0.93$&$+1.00$&$+0.07$&$-0.70$\\
8&$+0.31$&$-0.68$&$-0.82$&$-0.15$&$-0.18$&$-0.47$&$+0.11$&$+0.07$&$+1.00$&$+0.13$\\
9&$+0.76$&$-0.50$&$-0.40$&$-0.89$&$+0.47$&$-0.74$&$-0.67$&$-0.70$&$+0.13$&$+1.00$\\
\end{tabular}
\end{table}

\section{Determination of the effective leptonic weak mixing angle}
In order to determine the value of \stwlept that best describes the measured \afb distribution, predictions of \afb are produced using the \powhegbox program~\cite{powheg1,powheg2,powheg3}, using different configurations. 

The baseline prediction, hereafter referred to as `\powhegewnlo', takes NLO accuracy for both QCD and electroweak interactions~\cite{POWHEGewZ,POWHEGewZ2}, using the scheme described in Ref.~\cite{ewschemes}, that takes $G_\mu$, $m_Z$ and \stwlept as inputs. The events produced are then processed with \photos~\cite{Barberio:1990ms} for modelling of additional QED radiation and with \pythia8~\cite{Sjostrand:2007gs} for simulating the rest of the event.

Further predictions are produced to study modelling variations. Events are generated using the configuration described above but with the electroweak interactions simulated at LO accuracy; this is referred to as `\powhegewlo'. 
Predictions are also produced using an alternative calculation of the single-boson process in \powhegbox~\cite{POWHEGBoxV2SingleBoson} where QCD interactions are simulated at NLO accuracy and electroweak interactions are simulated at LO accuracy. For this prediction both additional QED radiation and additional
simulation of the rest of the event are performed using \pythia8. This configuration is labelled `\powhegplain'. These predictions are validated by producing an additional set of theoretical predictions using the $G_\mu$ input scheme~\cite{ewschemes} using both \powhegbox and \dyturbo~\cite{Camarda:2019zyx}. The two predicted \afb distributions show excellent agreement. 

The baseline description of the proton internal structure in all predictions uses the parton distributions from the central NNPDF3.1 PDF set at NLO~\cite{Ball:2017nwa}. Event weights are then used to recast the \powhegplain predictions to alternative parton distributions functions~\cite{Buckley:2014ana}. In this analysis predictions at NLO accuracy using the CT18~\cite{Hou:2019efy} and MSHT~\cite{Bailey:2020ooq} descriptions of the proton internal structure are also considered and treated equally to those from NNPDF3.1. These three descriptions all use broadly comparable global datasets and do not include the LHCb data studied here in their global fits. Other descriptions of the proton are also considered (NNPDF 4.0~\cite{NNPDF:2021njg}, CT18Z~\cite{Hou:2019efy}). 

In addition, events are generated using the \powhegplain configuration with variations in the QCD modelling. Events are generated with the factorisation and renormalisation scales varied by a factor of two around their baseline values in line with the seven-point variation approach~\cite{Hamilton:2013fea}, in order to assess the impact of missing higher-order
effects on the theoretical predictions. Events are also generated with two values of the strong coupling $\alpha_s$, 0.118 (the baseline) and 0.125. While this is a large variation with respect to the uncertainty on the world average value, this shift was observed to best describe the vector-boson \pt distribution in the LHCb measurement of the $W$-boson mass~\cite{LHCb-PAPER-2021-024}, and is again considered as a variation that mimics the effects of higher-order contributions in the predictions.

In order to determine the values of the weak mixing angle that best describe the data, predictions of \afb are made using events generated with different values of the weak mixing angle. Predictions for \afb at intermediate values are then found by interpolating between the generated base predictions. As a cross-check, the effect of including additional base predictions is also studied.

\begin{figure}
    \centering
    \includegraphics[width=0.7\textwidth]{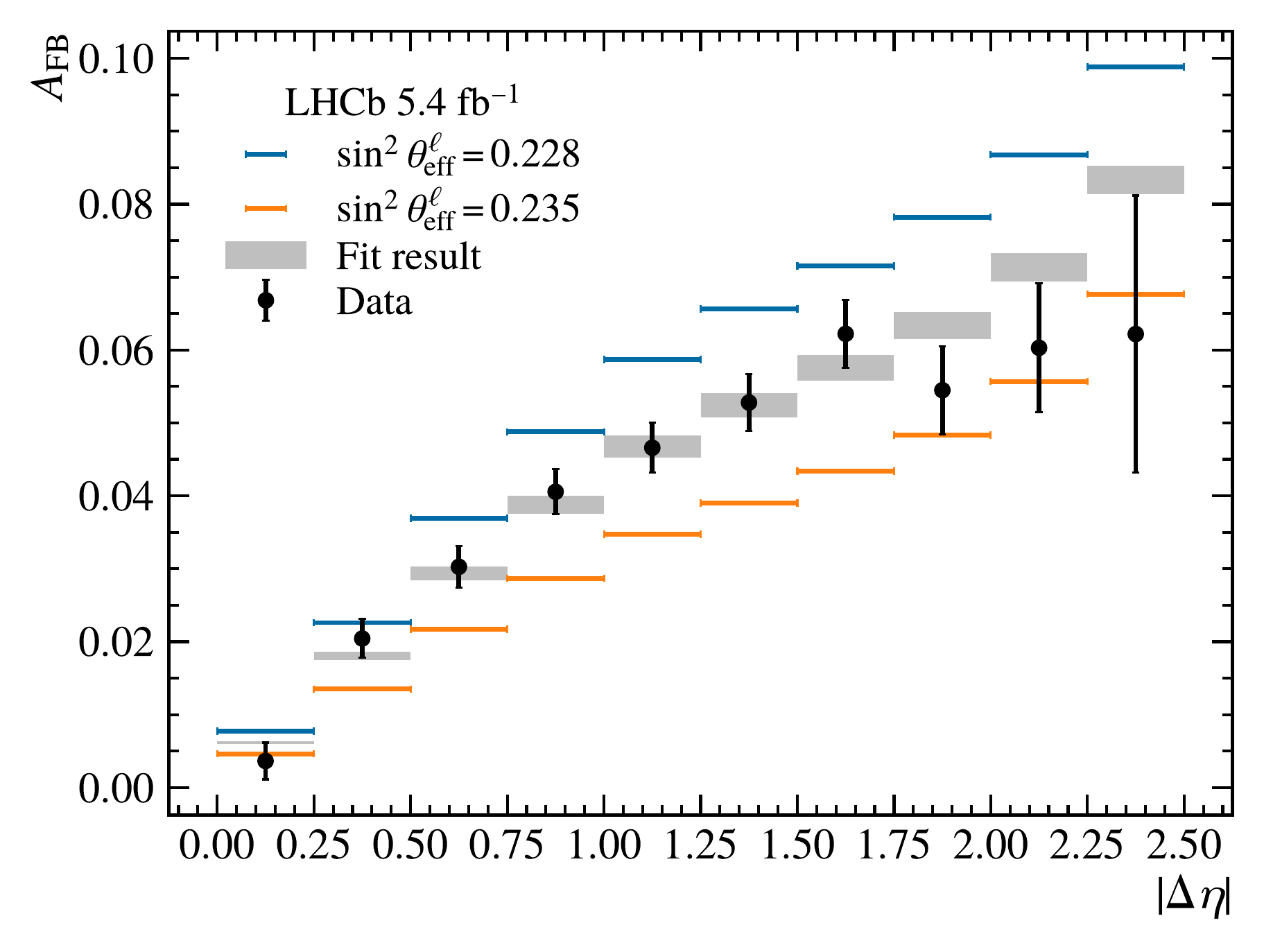}
    \caption{Measured \afb in ten intervals of \deltaeta, with the results of the \stwlept fit. The grey band shows the fit result and the associated statistical uncertainty.}
    \label{fig:afb_and_stw_fit}
\end{figure}

The analysis proceeds through a $\chi^2$ comparison of the measured \afb distribution to the theoretical predictions with different values of \stwlept, where the minimum of the $\chi^2$ comparison is used to determine the value of \stwlept, and the width of the $\chi^2$ parabola is used to determine the uncertainty. Figure~\ref{fig:afb_and_stw_fit} shows the measured \afb values compared to the predictions with two different \stwlept values and the baseline-fit result. 
The best fit point has a $\chi^2$ of $8.1$ for nine degrees of freedom (ndof), and results in
\begin{equation*}
\stwlept = 0.23148 \pm 0.00044 \pm 0.00005,
\end{equation*}
where the first uncertainty is statistical and the second results from propagating the 
systematic uncertainties on the \afb measurement. 

Several variations in the fit model are considered. Some of these variations are used to determine shifts to this result, while others set uncertainties or define cross-checks.

The default analysis uses two base templates for the \afb predictions at different values of \stwlept, with linear interpolation used to find predictions for \afb between these values. However, the impact of using a third base template is studied, applying cubic spline interpolation. A shift to the extracted result corresponding to the difference between these two approaches is applied, so that the final result is based on the cubic approach.\footnote{The application of this shift is equivalent to using three templates to find the central result, but only using two templates to evaluate uncertainties.} This provides a shift of $\NonLinearityShift$, consistent with the uncertainty associated with the number of generated events used to find the theoretical predictions. The resulting measurement of \stwlept is then found to be stable at the $1\times 10^{-5}$ level when the number of base templates is further increased to seven, confirming that the use of a small number of templates for the baseline result is reasonable.

An electroweak uncertainty of $7.4\times10^{-5}$ is assigned based on the difference between the result found using the \powhegewnlo and \powhegplain predictions. A cross-check is made using the \powhegewlo predictions, which are found to give results in agreement with \powhegplain, as expected. 

A QCD uncertainty is assigned based on changing the value of $\alpha_s$ used in the \powhegplain predictions to 
the value best describing
the data in the LHCb \PW-boson mass measurement~\cite{LHCb-PAPER-2021-024}. Since the change in the final result is smaller than the uncertainty on this shift from the number of generated events, the latter is assigned as the uncertainty on \stwlept, $5.8\times10^{-5}$. The number is consistent with an alternative estimate of the QCD uncertainty using predictions generated with the factorisation and renomalisation scales varied using the seven-point-variation method~\cite{Hamilton:2013fea}. 

The CT18, MSHT20 and NNPDF3.1 PDF parameterisations are treated equally. The final result quoted is therefore defined as the arithmetic average of the results from the three parameterisations. The impact of changing the PDF parameterisation is studied using \powhegplain events. The PDF uncertainties are determined for each PDF set using the prescription provided by each PDF-fitting group, by weighting the baseline events generated using the central NNPDF3.1 parameterisation. The CT18 uncertainties are divided by a factor $1.645$ in order to provide 68\% coverage.  It is found that changing from the baseline NNPDF3.1 result to the arithmetic average results in a shift of $-3\times 10^{-5}$. The PDF parameterisations are treated as fully correlated since they consider the same global data, and therefore the individual PDF uncertainties from the three parameterisations are averaged in order to set the overall PDF uncertainty on the measurement. The results from the different PDF sets used to define the central result are reported in Table~\ref{tab:stwlept_different_pdfs}. The impact of using other PDF sets is also studied and summarised in Table~\ref{tab:stwlept_extra_pdfs}. 

\begin{table}\centering
\caption{Fit results, using \powhegplain, for the PDF sets averaged in this analysis. The best-fit \stwlept values are listed, as are the PDF uncertainties, the shifts in the \stwlept values with respect to the first row and the fit $\chi^2/$ndof. 
The final row shows the shift that would be applied to the baseline result in order to emulate an arithmetic average of the three PDF sets, and the corresponding PDF uncertainty. The numbers presented in this table do not include the shift associated with changing from two base templates to three base templates. The PDF sets are labeled using the appropriate strings that fully define the set~\cite{lhapdf}.
}
\label{tab:stwlept_different_pdfs}
\begin{tabular}{lcccc}
PDF set & \stwlept & PDF uncertainty & Shift& Fit $\chi^2/$ndof \\
\hline 
NNPDF31\_nlo\_as0118  & $0.23155$ & $0.00023$ & -- & 8.4/9 \\
CT18NLO & $0.23165$ & $0.00022$ & $+0.00009$ & 8.4/9\\
MSHT20nlo\_as118 & $0.23137$ & $0.00017$ & $-0.00018$ & 8.2/9\\
\hline 
Arithmetic average & -- & $0.00021$ & $-0.00003$ & --\\
\end{tabular}
\end{table}

\begin{table}\centering
\caption{Fit results, using \powhegplain, for additional PDF sets not averaged for the final result. All other information given is the same as in Table~\ref{tab:stwlept_different_pdfs}. The shift is quoted relative to the NNPDF31\_nlo\_as0118 PDF set. 
}
\label{tab:stwlept_extra_pdfs}
\begin{tabular}{lcccc}
PDF set & \stwlept & PDF uncertainty & Shift& Fit $\chi^2/$ndof \\
\hline 
CT18ZNLO & $0.23147$ & $0.00019$ & $-0.00008$&8.4/9\\
NNPDF40\_nlo\_as\_01180 & $0.23142$ & $0.00022$ & $-0.00014$&8.6/9\\
\end{tabular}
\end{table}

\begin{figure}
    \centering
    \includegraphics[width=\textwidth]{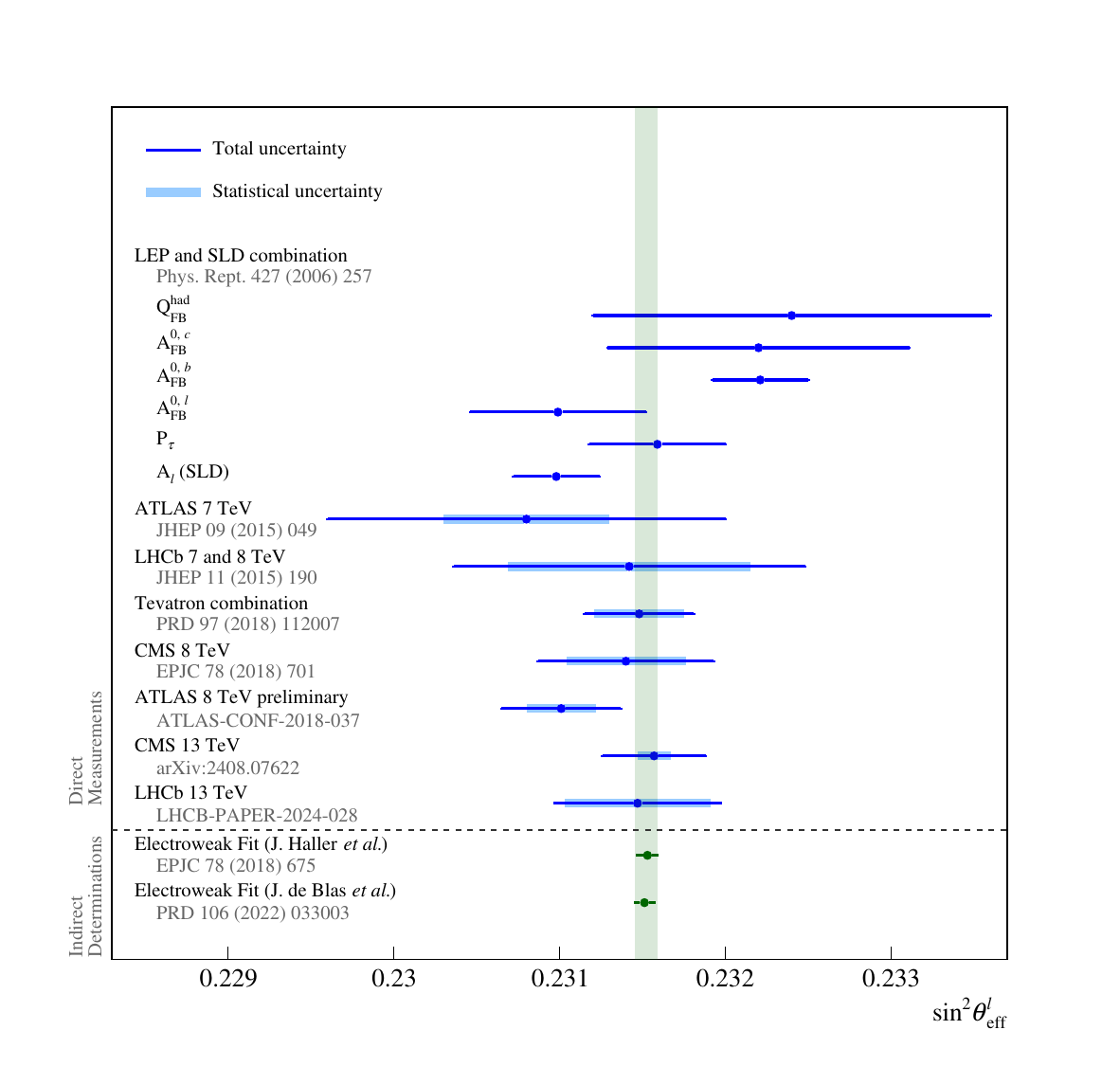}
    \caption{Direct measurements and indirect determinations of \stwlept. For the measurements from LEP and SLD only the total uncertainty is shown. The indirect determinations shown here include the LEP and SLD measurements as separate inputs while predicting the measurement at hadron colliders.}
    \label{fig:stwlept_summary}
\end{figure}

The studies using a larger number of base templates and averaging the results from the three different PDF parameterisations introduce a net shift of $-1\times10^{-5}$ relative to the baseline-fit result. Having applied this shift and included the theoretical uncertainties defined above, the final result is then 
\begin{equation*}
  \stwlept = \FinalResultValue \pm \FinalResultStatErr \pm \FinalResultExpErr \pm \FinalResultTheoryErr,
\end{equation*}
where the first uncertainty is statistical, the second is associated with systematic uncertainties on the \afb measurement, and the third is associated with theoretical uncertainties on the model used to determine the weak mixing angle. 
Figure~\ref{fig:stwlept_summary} compares this result with other measurements and with the Standard Model predictions. The LHCb measurement is in excellent agreement with previous measurements and with indirect determinations of the weak mixing angle from the global electroweak fit. It is also notable that while the theoretical uncertainty on the result is dominated by the PDF uncertainty, this uncertainty is also significantly smaller than the statistical uncertainty on the measurement. Consequently this analysis does not need to make use of profiling techniques to control and reduce the PDF uncertainty~\cite{Azzi:2019yne,Bodek:2015ljm}. However, such approaches are expected be of importance in future measurements, and projections are provided in Refs.~\cite{Azzi:2019yne,LHCb-PUB-2018-013}.

\section{Cross-checks}
Various cross-checks are performed to confirm the robustness of the data analysis. In these checks the baseline fit is performed. No shifts are applied to account for the change from two to three base templates, and no average is taken across the different PDF sets. Except where otherwise specified, results in this section are therefore reported using predictions based on \powhegewnlo and the NNPDF3.1 PDF parameterisation. In addition, no systematic uncertainties are considered when performing these checks.

Table~\ref{tab:cross_check_subsets} shows \stwlept fit results with the data divided into statistically independent subsets according to the year of data taking, the polarity of the magnet and the orientation of the decay with respect to the magnetic field, which is characterised by the angle $\phi_d$.\footnote{See, for example, Eq.~5 of Ref.~\cite{LHCb-DP-2023-001}.}
All three sets of results are self-consistent within their statistical uncertainties. 

\begin{table}\centering
\caption{Fit results with different subsets of the data.
For each subset, the first line is treated as the reference for the calculation of the pull. Each row has the same number of intervals and a ndof of 9.
}
\label{tab:cross_check_subsets}
\begin{tabular}{lccc}
Subset & \stwlept & Fit $\chi^2$/ndof & Pull \\
\hline
2016  & $0.23014 \pm 0.00082$ & $2.0/9$ & -- \\
2017  & $0.23155 \pm 0.00085$ & $13.4/9$ & $+1.2~\sigma$ \\
2018  & $0.23242 \pm 0.00077$ & $10.5/9$ & $+2.0~\sigma$ \\
\hline
Down polarity & $0.23087 \pm 0.00065$ & $8.2/9$ & -- \\
Up polarity & $0.23211 \pm 0.00065$ & $12.1/9$ & $1.4~\sigma$ \\
\hline 
$0 \leq \phi_d < \frac{\pi}{2}$ & $0.23136 \pm 0.00065$ & $10.1/9$ & -- \\
$\frac{\pi}{2} \leq \phi_d < \pi$ & $0.23161 \pm 0.00065$ & $6.5/9$ &  $+0.3~\sigma$\\
\end{tabular}
\end{table}

Table~\ref{tab:cross_check_delta_eta_intervals} presents \stwlept fit results with different numbers of intervals in \deltaeta, varying between one and ten. 
Compared to the result with a single interval, a relative improvement in the statistical precision of around $14\%$, as already discussed, is seen in the result with ten intervals.
The $\chi^2$ values are reasonable in all cases, and the shifts in the central values are small,
considering the statistical uncertainty on the shift between the result with one interval and those with multiple intervals.

\begin{table}\centering
\caption{Fit results with different numbers of \deltaeta intervals. The first row is the reference for the shift, and the uncertainties are statistical only.}
\label{tab:cross_check_delta_eta_intervals}
\begin{tabular}{cccc}
Number of intervals & \stwlept & Shift & Fit $\chi^2$/ndof \\
\hline 
1 & $0.23151  \pm 0.00050$ & -- & -- \\
4 & $0.23167 \pm 0.00045$ & $+0.00016$ & $3.1/3$\\
6 & $0.23145 \pm 0.00044$ & $-0.00004$ & $3.2/5$\\
8 & $0.23146 \pm 0.00044$ & $-0.00003$ & $11.7/7$\\
10 & $0.23148 \pm 0.00044$ & $-0.00003$ & $8.1/9$\\
\end{tabular}
\end{table}

To provide a cross-check using an alternative approach, the analysis is performed with a single \deltaeta interval but with seven intervals in the dimuon invariant mass.
Since the mass is measured with a resolution of $\mathcal{O}(1~\text{GeV})$, the migration is corrected for using 
iterative Bayesian unfolding~\cite{Dagostini2010}. Using the NNPDF3.1 PDF set this leads to a measurement of $\stwlept = 0.23130 \pm 0.00050$, with a $\chi^2$/ndof of $14.6/6$. The statistical precision of this check is poorer, by $14\%$, compared to our preferred approach of measuring \afb in intervals of \deltaeta.
The results remain stable when the number of intervals in the dimuon invariant mass is varied. The study is repeated and recast to additional PDF sets using the \powhegplain generator, with results reported in Table~\ref{tab:cross_check_mass_PDFs}. The MSHT20 and NNPDF4.0 parameterisations of the proton internal structure provide the best fit quality of those considered in this cross-check, though we note that no studies of systematic uncertainties are made for this cross-check, and that such studies could be expected to significantly impact this conclusion.

\begin{table}[b!]\centering
\caption{Fit results with different PDF sets for seven mass intervals. The shift of each result is reported relative to the first row.}
\label{tab:cross_check_mass_PDFs}
\begin{tabular}{lccc}
PDF set & \stwlept & Shift & Fit $\chi^2$/ndof \\
\hline 
NNPDF31\_nlo\_as0118  & 0.23133 & -- & 13.1/6 \\
CT18NLO & 0.23139 &  0.00006 & 19.8/6 \\
MSHT20nlo\_as118 & 0.23119 & -0.00015 & 10.8/6  \\
CT18ZNLO & 0.23126 &  -0.00007 & 17.1/6 \\
NNPDF40\_nlo\_as\_01180 & 0.23120 & -0.00014 & 9.4/6 \\
\end{tabular}
\end{table}

The following additional checks are also performed:
\begin{itemize}
\item In the \afb measurement, weights are assigned to the simulated signal events that shift the assumed \stwlept value. A shift corresponding to three times the uncertainty on the current world average causes a change in our measured \stwlept value below $2\times10^{-5}$, which is considered  negligible.
\item In addition to the results presented in Table~\ref{tab:cross_check_subsets}, measurements of \afb are performed with six orthogonal combinations of the year and magnet polarity. The resulting \stwlept results are statistically consistent.
\item Variations in the $\Upsilonres(1S)$ and \jpsi masses, within the uncertainties on their world averages, are propagated through the momentum calibrations; the effect on the measured \afb is negligibly small.
\item An alternative functional form is used in the momentum smearing which has a negligible effect on the results quoted.
\item The momentum smearing is calibrated solely using Z-boson events. The dimuon invariant mass distribution (as shown in Fig.~\ref{fig:simulation_control_plots}) exhibits a significantly better agreement between data and simulation with this change, with $\chi^2/\text{ndof}\sim 1$. However, because the analysis is insensitive to modelling of the muon momentum measurements, the measured value of the weak mixing angle  shifts by less than $1\times10^{-5}$ when this change is made.
\item Shifting the muon energies in the simulation, according to the uncertainties in the material budget of the detector, has a negligible effect on the results.
\end{itemize}

\section{Conclusion}

The effective leptonic weak mixing angle, \stwlept, is precisely predicted in the global electroweak fit. Direct measurements of this predicted quantity are sensitive to physics beyond the Standard Model.
A measurement of \stwlept is reported, based on $pp$ collision data at $\sqrt{s}=13$~TeV, recorded between 2016 and 2018 by the LHCb experiment and corresponding to an integrated luminosity of $5.4$~fb$^{-1}$.
The forward-backward asymmetry \afb in the $pp \to Z/\gamma^{*} \to \mu^+\mu^-$ process is measured in ten intervals of the difference of the muon pseudorapidities, within a fiducial region covering dimuon masses between $66$ and $116$~GeV, muon pseudorapidities between $2.0$ and $4.5$ and muon transverse momenta above $20$~GeV.
Comparing these forward-backward asymmetries with predictions at next-to-leading-order in the strong and electroweak couplings results in a determination of the effective leptonic weak mixing angle
\begin{equation*}
  \stwlept = \FinalResultValue \pm \FinalResultStatErr \pm \FinalResultExpErr \pm \FinalResultTheoryErr,
\end{equation*}
where the first uncertainty is statistical, the second is due to systematic uncertainties on the \afb measurement, and the third is due to theoretical uncertainties associated with the model used to determine the weak mixing angle. This result is based on an arithmetic average of results obtained using the CT18, MSHT20, and NNPDF3.1 parameterisations of the proton internal structure.
The result is consistent with other direct measurements and with predictions from the global electroweak fit, and improves on the precision of the previous LHCb determination by more than a factor two.

\section*{Acknowledgements}
%
%
\noindent We express our gratitude to our colleagues in the CERN
accelerator departments for the excellent performance of the LHC. We
thank the technical and administrative staff at the LHCb
institutes.
We acknowledge support from CERN and from the national agencies:
CAPES, CNPq, FAPERJ and FINEP (Brazil); 
MOST and NSFC (China); 
CNRS/IN2P3 (France); 
BMBF, DFG and MPG (Germany); 
INFN (Italy); 
NWO (Netherlands); 
MNiSW and NCN (Poland); 
MCID/IFA (Romania); 
MICIU and AEI (Spain);
SNSF and SER (Switzerland); 
NASU (Ukraine); 
STFC (United Kingdom); 
DOE NP and NSF (USA).
We acknowledge the computing resources that are provided by CERN, IN2P3
(France), KIT and DESY (Germany), INFN (Italy), SURF (Netherlands),
PIC (Spain), GridPP (United Kingdom), 
CSCS (Switzerland), IFIN-HH (Romania), CBPF (Brazil),
and Polish WLCG (Poland).
We are indebted to the communities behind the multiple open-source
software packages on which we depend.
Individual groups or members have received support from
ARC and ARDC (Australia);
Key Research Program of Frontier Sciences of CAS, CAS PIFI, CAS CCEPP, 
Fundamental Research Funds for the Central Universities, 
and Sci. \& Tech. Program of Guangzhou (China);
Minciencias (Colombia);
EPLANET, Marie Sk\l{}odowska-Curie Actions, ERC and NextGenerationEU (European Union);
A*MIDEX, ANR, IPhU and Labex P2IO, and R\'{e}gion Auvergne-Rh\^{o}ne-Alpes (France);
AvH Foundation (Germany);
ICSC (Italy); 
Severo Ochoa and Mar\'ia de Maeztu Units of Excellence, GVA, XuntaGal, GENCAT, InTalent-Inditex and Prog. ~Atracci\'on Talento CM (Spain);
SRC (Sweden);
the Leverhulme Trust, the Royal Society
 and UKRI (United Kingdom).

\addcontentsline{toc}{section}{References}
\bibliographystyle{LHCb}
\bibliography{main,standard,LHCb-PAPER,LHCb-CONF,LHCb-DP,LHCb-TDR}

\newpage
\centerline
{\large\bf LHCb collaboration}
\begin
{flushleft}
\small
R.~Aaij$^{37}$\lhcborcid{0000-0003-0533-1952},
A.S.W.~Abdelmotteleb$^{56}$\lhcborcid{0000-0001-7905-0542},
C.~Abellan~Beteta$^{50}$,
F.~Abudin{\'e}n$^{56}$\lhcborcid{0000-0002-6737-3528},
T.~Ackernley$^{60}$\lhcborcid{0000-0002-5951-3498},
A. A. ~Adefisoye$^{68}$\lhcborcid{0000-0003-2448-1550},
B.~Adeva$^{46}$\lhcborcid{0000-0001-9756-3712},
M.~Adinolfi$^{54}$\lhcborcid{0000-0002-1326-1264},
P.~Adlarson$^{81}$\lhcborcid{0000-0001-6280-3851},
C.~Agapopoulou$^{14}$\lhcborcid{0000-0002-2368-0147}
% ...

{\footnotesize \it

% ...
$^{14}$Universit{\'e} Paris-Saclay, CNRS/IN2P3, IJCLab, Orsay, France\\
% ...
$^{37}$Nikhef National Institute for Subatomic Physics, Amsterdam, Netherlands\\
% ...
$^{46}$Instituto Galego de F{\'\i}sica de Altas Enerx{\'\i}as (IGFAE), Universidade de Santiago de Compostela, Santiago de Compostela, Spain\\
% ...
$^{50}$Physik-Institut, Universit{\"a}t Z{\"u}rich, Z{\"u}rich, Switzerland\\
% ...
$^{54}$H.H. Wills Physics Laboratory, University of Bristol, Bristol, United Kingdom\\
% ...
$^{56}$Department of Physics, University of Warwick, Coventry, United Kingdom\\
% ...
$^{59}$School of Physics and Astronomy, University of Glasgow, Glasgow, United Kingdom\\
$^{60}$Oliver Lodge Laboratory, University of Liverpool, Liverpool, United Kingdom\\
% ...
$^{68}$Syracuse University, Syracuse, NY, United States\\
% ...
$^{81}$Department of Physics and Astronomy, Uppsala University, Uppsala, Sweden, associated to $^{59}$\\
% ...
}
\end{flushleft}

\end{document}